\documentclass{article}
\usepackage{spconf,amsmath,graphicx}

\usepackage{multirow}
\usepackage{booktabs}

\usepackage{hyperref}

\usepackage{xcolor}

\title{Self-supervised Speech Models 
 \\ for Word-Level Stuttered Speech Detection}
\name{Yi-Jen Shih, Zoi Gkalitsiou, Alexandros G. Dimakis, David Harwath}
\address{The University of Texas at Austin}

\begin{document}
%
\maketitle
\begin{abstract}
Clinical diagnosis of stuttering requires an assessment by a licensed speech-language pathologist. However, this process is time-consuming and requires clinicians with training and experience in stuttering and fluency disorders. Unfortunately, only a small percentage of speech-language pathologists report being comfortable working with individuals who stutter, which is inadequate to accommodate for the 80 million individuals who stutter worldwide. Developing machine learning models for detecting stuttered speech would enable universal and automated screening for stuttering, enabling speech pathologists to identify and follow up with patients who are most likely to be diagnosed with a stuttering speech disorder.
Previous research in this area has predominantly focused on utterance-level detection, which is not sufficient for clinical settings where word-level annotation of stuttering is the norm. 
In this study, we curated a stuttered speech dataset with word-level annotations and introduced a word-level stuttering speech detection model leveraging self-supervised speech models.
Our evaluation demonstrates that our model surpasses previous approaches in word-level stuttering speech detection. 
Additionally, we conducted an extensive ablation analysis of our method, providing insight into the most important aspects of adapting self-supervised speech models for stuttered speech detection.
\end{abstract}
\begin{keywords}
Self-supervised Speech model, Stuttering, Speech Pathology
\end{keywords}

\section{Introduction}
\label{sec:intro}
\begin{figure*}
\includegraphics[width=\textwidth,trim={0cm 0.5cm 2cm 0.2cm},clip]{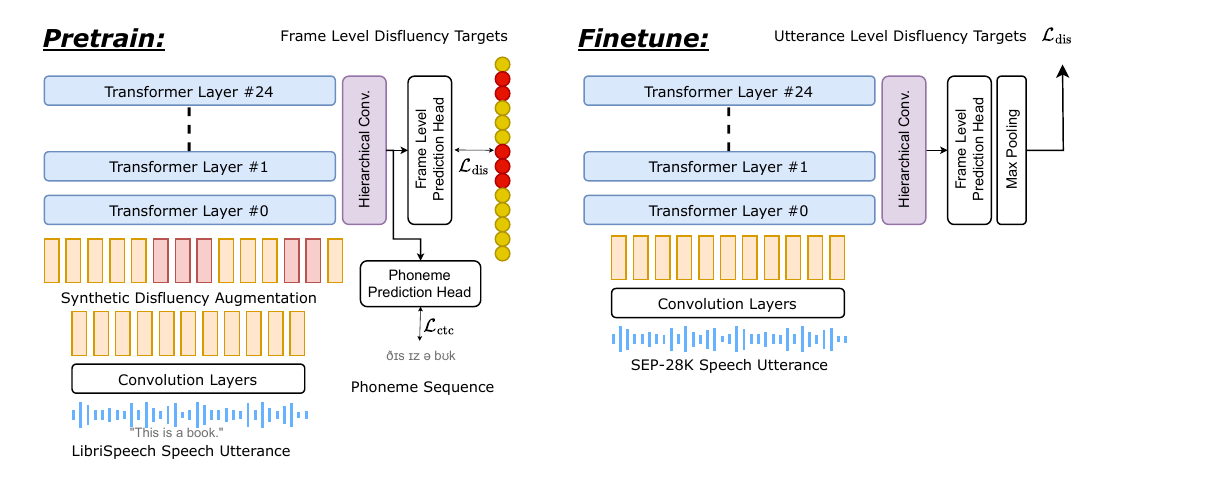}
\vspace{-20pt}
\caption{Overall framework diagram. The transformer layers and Convolution Layers are initialized from off-the-shelf WavLM Large and remain frozen during pretrain and finetune stage. ``Hierarchical Conv.'' stands for ``Hierarchical Convolution''. 
}
\vspace{-10pt}
\label{fig:model_arch}
\end{figure*}
Stuttering is a complex, multifactorial disorder characterized by atypical disruptions in the forward flow of speech~\cite{Smith_and_Weber}. It affects approximately 1 percent of the population and has a significant negative impact on all aspects of an individual’s life including social, educational, emotional, and vocational~\cite{craig_and_Tran,Koedoot_2011}.
A stuttering assessment is conducted by a licensed speech-language pathologist. It includes measurement of the impact of stuttering on the person’s quality of life as well as analysis of the individual’s speech to determine stuttering severity. Unfortunately, over the past decades, speech-language pathologists have consistently reported limited competence in working with individuals who stutter 
(e.g.,~\cite{Brisk_1997,Gabel_2013,kelly_1997,stLouis_1994}
, with less than 5 percent of licensed professionals in the United States reporting expertise working with individuals who stutter~\cite{Coalson_2016,kelly_1997}.
Another challenge is that the process of speech transcription and disfluency annotation is labor-intensive, as speech-language pathologists primarily transcribe and annotate speech disfluencies manually.  

A potential approach to address the above challenges is to use deep learning models to serve as an initial screening for stuttering, which could be deployed on edge devices such as smartphones or laptops. 
Early work~\cite{sheikhStutterNet,Kourkounakis2021FluentNetED}
has used simple deep learning models to detect stuttering behavior from speech utterances.
However, the main disadvantage is that the system is highly dependent on the scale of training data.
Unfortunately, the largest publicly available stuttering dataset is SEP-28K~\cite{lea_Sep28k}, which only contains 15.6 hours of utterance level labeled data, which is relatively small compared to standard datasets like LibriSpeech~\cite{librispeech}, which has a total 960 hours for training.

A promising recent approach in the speech community to deal with limited labeled data is Self-Supervised Learning (SSL).
Speech SSL models are first pretrained on a large amount of untranscribed speech and then finetuned on a smaller amount of transcribed/annotated speech for specific downstream tasks.
Due to the task-agnostic nature of the pretraining, speech SSL models demonstrate high generalizability across different speech processing tasks~\cite{yang21c_superb}.
Hence, speech SSL models are regarded as foundation models in many applications nowadays, such as Automatic Speech Recognition~\cite{baevski2021wav2vec-u,liu2022wav2vec-u2} and Speaker Verification~\cite{Peng_SSL_sv_2023}.
Due to the advantage of reducing the need for disfluency-labeled data, speech SSL models are a promising approach to tackling stuttered speech detection problems.
Specifically, prior works~\cite{Payal_iccasp_Siamese,bayerl22b_interspeech,bayerl23_interspeech} utilize Wav2vec 2.0~\cite{wav2vec2} as their model backbone for stutterred speech detection.

Despite progress in this field, previous work has mainly focused on utterance-level stuttering detection.
However, this is too coarse for clinical application as stuttering-like disfluencies are also present in typically fluent individuals who are not diagnosed as a Person who Stutters~(PWS).
Clinically, stuttering is diagnosed based on meeting a certain frequency threshold of stuttering-like disfluencies.  
Hence, we are interested in developing a more fine-grained stuttering detection model.
To our best knowledge, ~\cite{harvill22_interspeech} is the only previous work that investigates the task of frame-level stuttering detection.
However, they evaluated their model by removing the segments that are predicted as stuttering and conducted a user study on Amazon Mechanical Turk, which was not assessed on clinical data collected and annotated by speech-language pathologists (SLPs).

In our work, we first collect a set of speech recordings with stuttering events annotated by speech-language pathologists and use it as a word-level stuttering detection evaluation benchmark.
Then we propose a word-level stuttering detection model that utilizes WavLM~\cite{Chen2021WavLMLS} as our backbone.
Following~\cite{harvill22_interspeech}, we first pretrain on LibriSpeech with synthetic disfluency augmentations and then finetune our model on SEP-28K dataset.
We show that our model not only shows strong performance on utterance-level stuttering detection on SEP28K but also outperforms prior work on word-level stuttered speech benchmark by a large margin.
In addition, we study the effect of the utilization of Speech SSL models on word-level stuttering detection with extensive experiments.
In conclusion, our contribution can be summarized as the following:
\begin{enumerate}
    \item We are the first to focus on word-level stuttering detection, which is more closely aligned to clinical applications.
    \item We propose a word-level stuttering detection model that achieves state-of-the-art performance.
    \item We study the effect of utilizing Speech Self-supervised models on word-level stuttering detection with clinical data evaluation and extensive ablations.
\end{enumerate}
We open source our code on Github\footnote{\url{www.github.com/atosystem/SpeechSSLStutterDetect}}.

\section{Related Work}
\label{sec:related_work}
In the field of stuttered speech detection, there are several proposed datasets, such as UClass~\cite{Howell_uclass}, KSoF~\cite{bayerl_KSoFKasselState_2022}, and Sep-28K~\cite{lea_Sep28k}.
UClass is an unlabeled dataset while the others are labeled.
KSoF and Sep-28K share the same format.
Both of them consist of 3-second clips with utterance-level annotations.
Specifically, Sep-28K also includes a subset of utterance level annotation from Fluency Bank.
Not only the size of the dataset are relatively small, but they only provide 
utterance-level labels,
which we believe is one of the main obstacles in this field.

Several recent works have used Speech SSL models in stuttering detection.
In ~\cite{bayerl22b_interspeech}, they adopted Wav2vec 2.0 as their backbone and trained different models using different upstream layers of Wav2vec 2.0 for stuttering detection.
They also proposed to employ gender prediction as an auxiliary task for enhancing performance.
In ~\cite{bayerl23_interspeech}, they pointed out that stuttering types do not happen independently, so they framed the detection problem as a multi-label classification problem.
In ~\cite{Payal_iccasp_Siamese}, they improved the quality of Wav2vec 2.0 embeddings for stuttered speech detection using Siamese network and contrastive training. 
Despite these impressive results, their models predict at the utterance level, which is insufficient for clinical usage.
Furthermore, the optimal strategies for utilizing Speech SSL models for stuttering detection are still not fully understood.
Several prior works also incorporate Speech SSL models into their model design, such as for detecting dysarthria~\cite{Farhad_Dysarthria} and aphasia~\cite{Lian2023UnconstrainedDM,Lian2024TowardsHS}, but they focus on other types of speech disorders and are not directly applicable to our task.

To the best of our knowledge, ~\cite{harvill22_interspeech} is the only study that focused on fine-grained stuttering detection, more specifically frame-level stuttering detection.
Hence, we chose it as our baseline.
It is first pretrained on LibriSpeech with synthetic disfluencies augmented during training with frame-level supervision.
Then, they added a max pooling layer to finetune on utterance-level SEP-28K dataset.

\section{Method}
\label{sec:method}
\subsection{Model Architecture}
Following prior work using Wav2vec 2.0 on stuttered speech detection, we chose another Speech SSL model,  WavLM Large~\cite{Chen2021WavLMLS} as our backbone since it significantly outperforms Wav2vec 2.0 on most SUPERB~\cite{yang21c_superb} benchmark tasks.
Our model structure is shown in Figure~\ref{fig:model_arch}.
Furthermore, we adopt the recent Hierarchical Convolution Interface~(HConv.) as our method for utilizing the WavLM backbone according to recent work~\cite{shih_interface}, which was shown to improve overtaking a learnable weighted sum of layer activations.
As pointed out in~\cite{bayerl22b_interspeech}, the information for stuttering detection exists in multiple layers, so we decided to use HConv. due to its ability to aggregate information across multiple layers.
After that, we add a few non-linear layers as the prediction head to classify individual frames as stuttered or non-stuttered.

In~\cite{bayerl22b_interspeech}, it was shown that phoneme recognition and stuttering detection tasks utilize the same set of layers in the upstream Speech SSL model.
Motivated by this, we decided to add an auxiliary Connectionist Temporal Classification~(CTC) Loss to predict the phoneme sequences.
The CTC loss is only applied in the pretraining stage where we have the ground truth phoneme sequences.

We roughly follow the training pipeline in~\cite{harvill22_interspeech}, our framework consists of two stages: Pretrain on synthetic augmentation of LibriSpeech 360 and Finetune on SEP-28K.

\noindent
\textbf{Pretrain on LibriSpeech}:
Different from previous works, we add our synthetic augmentations between the convolution layers and the transformer layers of WavLM rather than adding augmentation directly on the input waveform.
We keep the augmentation types the same as~\cite{harvill22_interspeech}, which includes artificial prolongation, and word/sound repetition.
By doing so, we significantly save on computation and speed up the online augmentation.
Suppose the outputs of the prediction head for each input utterance are $\mathbf{o}=\left[ \mathbf{o}_1,\mathbf{o}_2,\dots \mathbf{o}_T \right]$, where $\mathbf{o}_i=\left[ o^p_i, o^n_i\right]$, corresponding to the probability of positive and negative for frame $i$ respectively.
The labels are $\mathbf{l}=\left[ \mathbf{l}_1,\mathbf{l}_2,\dots \mathbf{l}_T \right]$.
The loss is calculated as a cross entropy between the prediction head outputs and the label sequence.
\begin{align}
    \mathcal{L}_{\mathrm{dis}} = \frac{1}{T}\sum_{i=1}^{T}{
    -o^p_i \log{l^p_i}
    -o^n_i \log{l^n_i}},
\end{align}
$\mathcal{L}_{\mathrm{ctc}}$ is calculated between the frame level output of Phoneme Prediction head and the ground truth phoneme sequence in LibriSpeech~(which is obtained from a force aligner).
The overall pretrain loss is:
\begin{align}
    \mathcal{L}_{\mathrm{pretrain}} = \mathcal{L}_{\mathrm{dis}} + w_{\mathrm{ctc}} \cdot \mathcal{L}_{\mathrm{ctc}},
\end{align} where $w_{\mathrm{ctc}}$ is a scalar hyperparameter.

\noindent
\textbf{Finetune on SEP-28K}:
For the fine-tuning stage, since we do not have frame-level labels for SEP-28K, we simply take the timestep with the largest positive class prediction to calculate cross entropy with the utterance level target in SEP-28K, which is same as~\cite{harvill22_interspeech}. Formally,
\begin{align}
   t'= \mathrm{argmax}_{1\leq i \leq T}{~o^p_i},
\end{align}
\begin{align}
   \mathcal{L}_{\mathrm{dis}} = 
    -o^p_{t'} \log{l^p_{t'}}
    -o^n_{t'} \log{l^n_{t'}}.
\end{align}


\subsection{Implementation Details}
\textbf{Model and training details:}
The scalar hyperparameter $w_{\mathrm{ctc}}$ is set to $0.3$ empirically to balance the magnitude between the two losses.
For the finetuning stage, we feed the entire audio without augmentation and calculate a global level loss with its corresponding label.
In both stages, the HConv. interface and the stuttering detection head are trainable.
We use Adam optimizer with a learning rate of $1e-4$.
We select the checkpoint with the lowest validation loss for the pretraining stage and for the finetuneing stage, we select the one with the highest F1 score on the validation set.

\noindent
\textbf{Dataset:}
For pretraining, we use LibriSpeech 360hr and for SEP-28K, we follow the split specified in~\cite{harvill22_interspeech} for SEP-28K as they make the dataset balanced across positive and negative classes.

\section{Word Level Stuttering Speech Dataset}
\label{sec:dataset}
To make our system clinically applicable, students in speech-language pathology, trained in fluency disorders and stuttering, transcribed and annotated all the speech samples using CHAT, a transcription program that is
part of CLAN and the Talk-Bank initiative~\cite{Macwhinney_2000}.
We used a set of codes developed to be used with
CHAT~\cite{Ratner_Brundage_2020}
to annotate stuttering disfluencies and typical disfluencies.
Our dataset comes from two sources:

\noindent\textbf{FluencyBank:} This dataset included interview data from 36 adult individuals who stutter from the 
FluencyBank~\cite{Ratner_2018}
English Voices-AWS Corpus. 

\noindent\textbf{Bilinguals Speech:} 
This dataset included narrative samples produced in English by 62 bilingual adults, 6 of whom stuttered. 
Participants were asked to generate a story based on a wordless picture book. 
In our evaluation, we have two partitions for this dataset: \textit{Stuttering Bilinguals Speech}, \textit{Non-stuttering Bilinguals Speech}.

To conduct a word-level stuttering evaluation on the CHAT annotations, we need timestamp information.
To this end, we first leverage Whisper Large~\cite{pmlr-v202-radford23a_whisper} to transcribe the corresponding audio file and get the transcripts as well as the timestamps for each word.
Then, we employ Dynamic Time Warping(DTW) to align the Whisper ASR transcripts with the human transcriptions in CHAT files and slice the audio into multiple utterances according to the lines in CHAT files.
For each aligned word between the Whisper ASR transcript and CHAT file transcript, we label the word as stutter if there is any stuttering annotation in its CHAT annotation, which includes: Prolongation, Epenthesis, Broken Word, Block, Sound/Syllable Repetition.
We add a little margin~(less than 0.2 sec) on the Whisper timestamps of each word due to the fact that Whisper truncates the audio segment where Block manifests. 
Notice that we only evaluate those words with exact alignment in DTW since we do not have word-level timestamps for unaligned words.
We end up obtaining about 14.43 min of \textit{Non-stuttering Bilinguals Speech}, 8.28 min of \textit{Stuttering Bilinguals Speech}, 45.69  min of \textit{FluencyBank} and a total of 1.14 hr recordings.

\section{Evaluation}
\label{sec:evaluation}
\begin{table*}
\vspace{-10pt}
\centering
\begin{tabular}{lcccccccc}
\toprule
\multirow{2}{*}{Model} & \multirow{2}{*}{SEP-28K Test}  & \multicolumn{6}{c}{SEP-28K - Fluency Bank} \\
\cmidrule{3-8}
 &   & All	& Blocks & Interjections & Prolongation & Sound Rep. & Word Rep.\\
\midrule
Baseline~\cite{bayerl22b_interspeech}$^\dagger$ & -- &
 --   & 0.33 &	\textbf{0.84} &	0.60 &	0.60 &	0.43 \\
Baseline~\cite{bayerl23_interspeech}$^\dagger$ & -- &
 --     & 0.17 & \underline{0.82} & 0.60 & 0.63 & 0.47 \\
\midrule
Baseline~\cite{harvill22_interspeech} & 0.700 &
0.68 &	0.34 &	0.62 &	0.28 &	0.39 &	0.39 \\

WavLMLg + HConv. & \underline{0.800} &	\textbf{0.85} &	\textbf{0.63} &	0.59 &	\textbf{0.63} &	\textbf{0.75} &	\textbf{0.70} \\
WavLMLg + HConv. + CTC & \textbf{0.803} &	\underline{0.84} &	\underline{0.62} &	0.57 &	\underline{0.61} &	\underline{0.74} &	\underline{0.67} \\

\midrule
 & \multicolumn{7}{c}{\textit{Using Weighted Sum~(WS)}} \\
 \cmidrule(lr){2-8}
WavLMLg + WS & 0.789 &	0.82 &	0.58 &	0.63 &	0.54 &	0.70 &	0.63 \\
WavLMLg + WS + CTC & 0.790 &	0.82 &	0.59 &	0.59 &	0.55 &	0.69 &	0.62 \\

\midrule
 & \multicolumn{7}{c}{\textit{Different Finetune Dataset Size using ``WavLMLg + HConv. + CTC''}} \\
\cmidrule(lr){2-8}
Data Ratio: $0\%$ &
0.685 &	0.70 &	0.36 &	0.59 &	0.31 &	0.43 &	0.37 \\
Data Ratio: $25\%$ &
0.774 &	0.82 &	0.62 &	0.58 &	0.61 &	0.77 &	0.72 \\
Data Ratio: $50\%$ &
0.798 &	0.84 &	0.65 &	0.57 &	0.64 &	0.76 &	0.71 \\
Data Ratio: $75\%$ &
0.793 &	0.83 &	0.66 &	0.56 &	0.65 &	0.79 &	0.74 \\
Data Ratio: $100\%$ & 0.803 &	0.84 &	0.62 &	0.57 &	0.61 & 0.74 &	0.67 \\

\bottomrule
\end{tabular}
\caption{Utterance Level Stuttered Speech Detection Evaluation on SEP-28K and Fluency Bank. The numbers are F1 scores.
Notice that for models with $^\dagger$, they are trained to distinguish different stuttering types while training. Also the numbers are directly reported from their paper.}
\label{tab:Utt_level}
\vspace{-5pt}
\end{table*}

While we aim to evaluate the performance of our model on word-level stuttered speech detection,  we also want to compare it with previous works that studied utterance-level stuttering detection.
Hence, we evaluate under both the utterance-level and word-level stuttering detection conditions.
For the former, we evaluate on SEP-28K testing set and also report the F1 score for each stuttering type on Fluency Bank (a subset of SEP-28K).
For the latter, we evaluate models on our word-level stuttering dataset and report F1, precision, and recall on each partition separately.
Additionally, we report the F1 score and Average precision on the entire dataset.
We choose to report F1 score instead of other metrics such as accuracy due to the fact that the data is highly imbalanced~(There are many more non-stuttered speech segments than stuttered segments).
We sweep the detection threshold from 0 to 1 with a step of 0.05 to find the optimal threshold for all settings.
Throughout our evaluation, we found out that the word-level dataset is very sensitive to the threshold selection.
Hence, we also report Average Precision on the entire word-level stuttered speech dataset.
F1 score on the one hand is more intuitive but sensitive to threshold selection especially on this dataset, while Average Precision is a threshold-independent metric that evaluates the classification model under all possible thresholds.
With these two evaluation benchmarks, we further conduct ablation experiments on the effect of CTC loss, the SSL interface/model selection, and the size of the finetuning dataset.
Finally, we show some qualitative examples of our model's outputs.

\vspace{-10pt}
\subsection{Utterance Level stuttered speech detection}
Since our model is only trained to detect stuttering events and not to distinguish between different types of stuttering, we report the type-specific stuttering detection F1 score by selecting different subsets of the testing dataset.
For example, to test the F1 score for blocks on SEP-28K - Fluency Bank, we filter the testing set to include only utterances with blocks and utterances without stuttering.

As shown in Table~\ref{tab:Utt_level}, overall, our model surpasses the baseline systems.
Specifically, \cite{bayerl22b_interspeech,bayerl23_interspeech} and our model outperforms ~\cite{harvill22_interspeech} by a large margin, indicating the benefit of using Speech SSL models and probably the transformer architecture.
Interestingly, our model beats \cite{bayerl22b_interspeech,bayerl23_interspeech} in general, except for interjections.
This could potentially be improved by adding synthesized sound interjections in the pretraining stage.

\vspace{-10pt}
\subsection{Word Level stuttered speech detection}
\begin{table*}
\centering
\begin{tabular}{lccccccccccc}
\toprule
\multirow{2}{*}{Model} & \multicolumn{2}{c}{All} &  \multicolumn{3}{c}{Stut. Blngl. Speech} & \multicolumn{3}{c}{Non-stut. Blngl. Speech} &  \multicolumn{3}{c}{Stut. Fluency Bank} \\
\cmidrule(lr){2-3}
\cmidrule(lr){4-6}
\cmidrule(lr){7-9}
\cmidrule(lr){10-12}
   & F1 & Avg.~P & 
 F1	& P & R & 
 F1	& P & R & 
 F1	& P & R \\
\midrule
Baseline~\cite{harvill22_interspeech}& 0.411 & 0.864 &
0.389 &	0.360 &	0.423 &	
0.342 &	0.245 &	0.565 &	
0.440 &	0.315 &	0.728 \\

WavLMLg + HConv. & 
	0.536 & 0.899  &	0.484 &	0.364 &	0.723 &	0.496 &	0.391 &	0.680 &	0.585 &	0.496 &	0.715 \\

WavLMLg + HConv.+ CTC &
	\textbf{0.554} & \textbf{0.927}  &	\textbf{0.490} &	0.375 &	0.708 &	\textbf{0.518} &	0.439 &	0.633 &	0.591 &	0.508 &	0.708 \\

\midrule
 & \multicolumn{10}{c}{\textit{Using Weighted Sum~(WS)}} \\
\cmidrule(lr){2-12}
WavLMLg + WS. &
	0.536 & 0.898 &	0.425 &	0.356 &	0.529 &	0.425 &	0.356 &	0.529 &	0.594 &	0.507 &	0.718 \\

WavLMLg + WS. + CTC &
	0.540 & 0.888 &	0.460 &	0.336 &	0.730 &	0.403 &	0.344 &	0.486 &	\textbf{0.603} &	0.514 &	0.729 \\


\midrule
 & \multicolumn{10}{c}{\textit{Select Single WavLM Large Layer}} \\
\cmidrule(lr){2-12}

WavLMLg Layer~0 & 0.439 & 0.864 &	0.340 &	0.214 &	0.825 &	0.279 &	0.192 &	0.507 & 0.519 & 0.415 & 0.694 \\
WavLMLg Layer~4 & 0.479 & 0.876 &	0.403 &	0.292 &	0.650 &	0.332 &	0.256 &	0.471 & 0.539 & 0.434 & 0.709 \\
WavLMLg Layer~8 & 0.499 & 0.875 &	0.442 &	0.389 &	0.511 &	0.424 &	0.341 &	0.561 & 0.541 & 0.476 & 0.628 \\
WavLMLg Layer~12 & 0.522 & 0.891 &	0.458 &	0.331 &	0.745 &	0.417 &	0.334 &	0.554 & 0.581 & 0.493 & 0.707 \\
WavLMLg Layer~16 & 0.548 & 0.910 &	0.480 &	0.418 &	0.562 &	0.513 &	0.462 &	0.576 & 0.589 & 0.513 & 0.692 \\
WavLMLg Layer~20 & 0.552 & 0.903 &	0.486 &	0.378 &	0.679 &	0.488 &	0.415 &	0.594 & 0.592 & 0.521 & 0.684 \\
WavLMLg Layer~22 &	0.550 &  0.912 &	0.488 &	0.361 &	0.752 &	0.484 &	0.391 &	0.633 &	0.590 &	0.522 &	0.679 \\

WavLMLg Layer~24 & 0.504 & 0.901 &	0.464 &	0.366 &	0.635 &	0.429 &	0.345 &	0.568 & 0.546 & 0.470 & 0.651 \\


\midrule
 & \multicolumn{10}{c}{\textit{Different Finetune Dataset Size using ``WavLMLg + HConv. + CTC''}} \\
\cmidrule(lr){2-12}
Data Ratio: $0\%$  &
 0.469 & 0.872 & 0.377 & 0.299 & 0.511 & 0.305 & 0.259 & 0.371 & 0.545 & 0.457 & 0.674 \\
Data Ratio: $25\%$ &
  0.530 & 0.892 & 0.471 &  0.405 & 0.562 & 0.362 & 0.452 & 0.302 &	0.579 &	0.487 &	0.713 \\
Data Ratio: $50\%$ &
0.533 & 0.904 &	0.470 &	0.338 &	0.774 &	0.493 &	0.413 &	0.612 &	0.579 &	0.487 &	0.713 \\
Data Ratio: $75\%$ &
0.516 & 0.888 &	0.460 &	0.368 &	0.613 &	0.386 &	0.417 &	0.360 &	0.557 &	0.472 &	0.678 \\
Data Ratio: $100\%$ &
0.554 & 0.927  &	0.490 &	0.375 &	0.708 &	0.518 &	0.439 &	0.633 &	0.591 &	0.508 &	0.708 \\


\bottomrule
\end{tabular}
\caption{Word Level Stuttered Speech Detection Evaluation. ``WavLMLg'' stands for WavLM Large.
HConv. indicates Hierarchical Convolution Interface and WS means weighted sum interface. For the metrics, ``Avg.~P'' stands for Average Precision while ``P'' and ``R'' indicate precision and recall respectively.
``Stut. Blngl. Speech'' and ``Non-stut. Blngl. Speech'' stands for Stuttering Bilinguals Speech and Non-stuttering Bilinguals Speech.
}
\label{tab:Word_level}
\vspace{-10pt}
\end{table*}

In Table~\ref{tab:Word_level}, overall ``WavLMLg + HConv. + CTC'' outperforms the baseline model and other variations.
In terms of different partitions, Fluency Bank has the highest F1 scores regardless of method.
The reason might be the difference between native speakers and bilingual speakers.
For Bilinguals Speech, stuttering bilinguals tend to have higher F1 scores compared to non-stuttering bilinguals.
The reason is that our model tends to have a high recall and a low precision in most cases, which is also true for the baseline model.
Compared to the baseline method, our model improves more on recall than precision.
One avenue for future work would be to incorporate an improved prior over where stuttered speech is likely to occur within an utterance or a better way to condition the model on different patient demographics.

\vspace{-10pt}
\subsection{The effect of CTC loss and SSL interface selection}
To study the effect of CTC loss and HConv. Interface we design several ablation experiments.
For the utilization of WavLM, we design models using weighted sum~(which is a set of learnable weights that is multiplied layer-wisely on WavLM Large hidden layers and sum up) or simply select a single layer.
Due to the extensive computing, we select every 4 layers of WavLM Large for our experiment.
For HConv. and Weighted Sum, we also try adding CTC loss or not.

\begin{figure}
\centering
\includegraphics[width=0.5\textwidth,trim={0.3cm 0.3cm 0.3cm 0.4cm},clip]{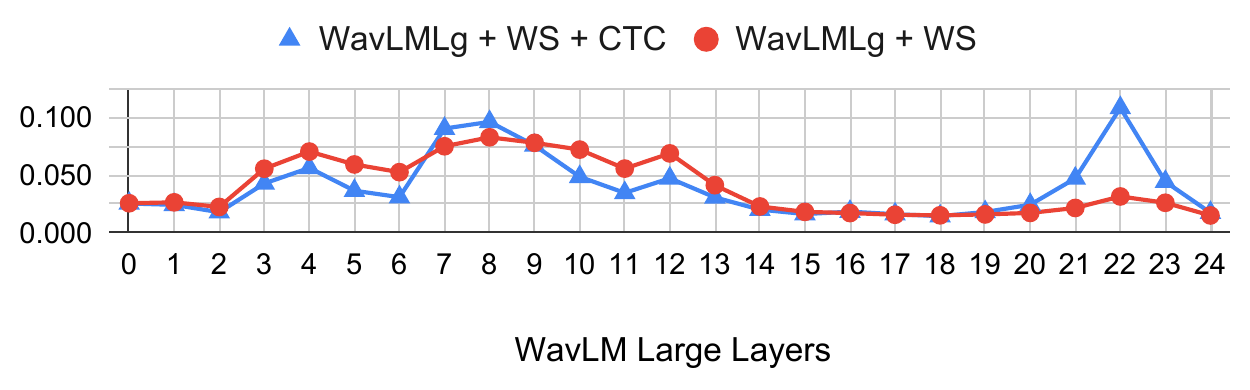}
\caption{The learned weights distribution of weighted sum interface in WavLMLg + WS and WavLMLg + WS + CTC.}
\label{fig:ws_dist}
\vspace{-10pt}
\end{figure}
\begin{figure*}[h]
\centering
\includegraphics[width=0.9\textwidth,trim={0.2cm 38cm 7cm 0.9cm},clip]{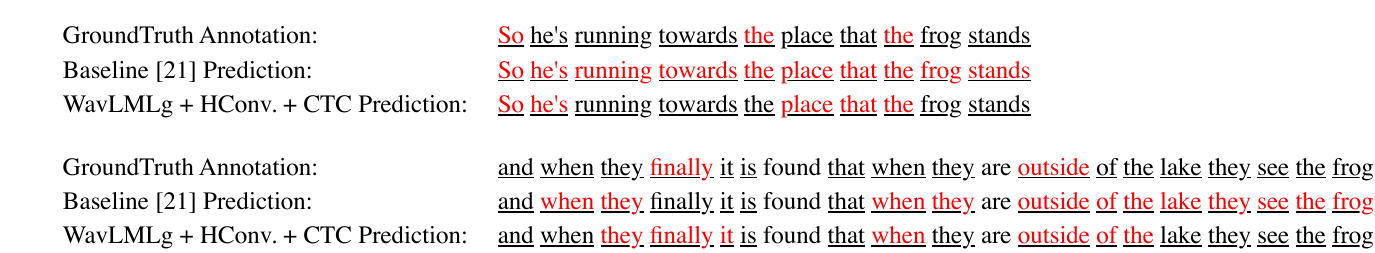}
\caption{Examples of our model and baseline model on word-level stuttering speech detection. Only the words with underscores are considered when evaluating. Words in red indicate that it is categorized as stuttered speech.
}
\label{fig:qualititave_examples}
\vspace{-10pt}
\end{figure*}
On the utterance level (See Table~\ref{tab:Word_level}), we see that with the help of CTC loss, the performance with both HConv. and  Weighted Sum layer aggregation strategies improved.
Interestingly, without CTC loss, both interfaces perform roughly the same overall.
We hypothesize that HConv. could benefit more from the guidance of CTC loss in terms of word-level stuttered speech detection.
The same trend can also be observed on the word-level stuttering speech dataset in Table~\ref{tab:Word_level}.
Hence, we further visualize the learned weights of the weighted sum in ``WavLMLg + WS'' and ``WavLMLg + WS + CTC'' in Figure~\ref{fig:ws_dist}.
With the help of CTC loss, the weight distribution becomes more concentrated on the peaks, indicating that CTC loss is helping the model to learn a better way of utilizing upstream model features.

We also report the performance of using a single layer from WavLM-Large in Table~\ref{tab:Word_level}.
Besides experiments with every four layers of WavLM-Large, we also include an experiment using the 22nd layer of WavLM-Large according to the layer with the highest learned weight in Figure~\ref{fig:ws_dist}.
Interestingly, layers 8 and 22 have the highest learned weights, but overall the best performance falls roughly on the upper half of WavLM-Large layers~(from layer 12 to 24).
We note that ``WavLMLg Layer 22'' has an overall F1 score close to our best model ``WavLMLg + HConv.+ CTC'', but falls behind in terms of average precision, indicating that the HConv. interface is an effective approach to utilizing WavLM-Large on word-level stuttered speech detection.

\vspace{-14pt}
\subsection{The effect of different Speech SSL models}
\begin{table}
\centering
\small
\begin{tabular}{lcc}
\toprule
Model & SEP-28K Test & Word Level Stuttered 
\\
\midrule

Baseline~\cite{harvill22_interspeech} & 0.700 &	0.411 \\
WavLM Large & \textbf{0.803} &	\textbf{0.554} \\
WavLM Base & 0.747 &	0.484 \\
Wav2vec2 Large & 0.760 &	0.524 \\
Wav2vec2 Base & 0.753 &	0.483 \\
Data2vec Large & 0.751 &	0.549 \\
Data2vec Base & 0.745 &	0.530 \\

\bottomrule
\end{tabular}
\caption{The F1 score for using different Speech SSL Models. All models are trained using Hierarchical Convolution and CTC loss.}
\label{tab:different_ssl}
\vspace{-16pt}
\end{table}
\vspace{-4pt}
In addition to the interface selection, we also tried using different sizes and types of upstream Speech SSL models.
As shown in Table~\ref{tab:different_ssl}, overall, the Speech SSL models outperform the baseline model, indicating the effectiveness of self-supervised training.
Furthermore, Large size models surpass Base size models in all cases.
Last but not least, WavLM performs the best compared to Wav2vec2~\cite{wav2vec2} and Data2vec~\cite{baevski22a_data2vec}.
From our results, we believe that WavLM is a more suitable choice for stuttering detection and that previous works~\cite{Payal_iccasp_Siamese,bayerl22b_interspeech,bayerl23_interspeech} might also benefit if they change their backbone from Wav2vec2 to WavLM.

\vspace{-14pt}
\subsection{The effect of fine-tuning dataset size}
\vspace{-4pt}
To test our model on the effect of fine-tuning dataset size, we tried using different proportions of our fine-tuning data: $0\%,25\%,50\%,75\%,100\%$.
Notice that we keep the ratio of positive and negative examples in the training set the same while using different proportions of the data.
We report both utterance and word level results on Table~\ref{tab:Utt_level} and Table~\ref{tab:Word_level}.
On both datasets, we observe a large improvement from 0$\%$ to 50$\%$.
Surprisingly, there is a little drop in the performance when using 75$\%$ of the data, but there's a large leap when using 100 $\%$ of the data.
Overall, this again corroborates the problem of data scarcity in stuttered speech detection and shows our model's scalability.

\vspace{-14pt}
\subsection{Qualitative results of Stuttering Detection}
\vspace{-4pt}
To visualize our detection results, we show some of the examples in our word-level stuttered speech detection dataset.
As shown in Figure~\ref{fig:qualititave_examples}, both the baseline and our model tend to predict stuttering events more frequently than the speech pathologist.
However, compared with the baseline model, our model's predictions tend to be of higher precision.



\vspace{-12pt}
\section{Conclusions}
\label{sec:conclusion}
\vspace{-5pt}
In this paper, we shed light onto a more fine-grained evaluation of stuttered speech detection which is also closer to clinical usage.
We curated a clinical word-level stuttering speech dataset and further proposed a word-level stuttered speech detection model which exhibits significant improvements compared to prior work.
Additionally, through our comprehensive ablation studies, we investigated the utilization of Speech SSL models on stuttered speech detection and demonstrated the potential scaling of our method.

For future work, we think that increasing the dataset size and diversity would bring additional improvements for this field.
Finally, we would also like to incorporate word-level stuttering type classification in our model, to better assist speech pathologists for screening and diagnosis. 

\vspace{-14pt}
\section{Acknowledgement}
\label{sec:ack}
\vspace{-10pt}
This research has been supported by NSF Grants AF 1901292, CNS 2148141, IFML CCF 2019844 and research gifts by Cisco, WNCG IAP and the  UT Austin Machine Learning Lab (MLL).



\bibliographystyle{IEEEbib}
\bibliography{strings,refs}

\begin{thebibliography}{10}

\bibitem{Smith_and_Weber}
Anne Smith and Christine Weber,
\newblock ``How stuttering develops: The multifactorial dynamic pathways theory,''
\newblock {\em Journal of speech, language, and hearing research : JSLHR}, vol. 60, pp. 1--23, 08 2017.

\bibitem{craig_and_Tran}
Ashley Craig, Elaine Blumgart, and Yvonne Tran,
\newblock ``The impact of stuttering on the quality of life in adults who stutter,''
\newblock {\em Journal of Fluency Disorders}, vol. 34, no. 2, pp. 61--71, 2009.

\bibitem{Koedoot_2011}
Caroline Koedoot, Clazien Bouwmans, Marie-Christine Franken, and Elly Stolk,
\newblock ``Quality of life in adults who stutter,''
\newblock {\em Journal of Communication Disorders}, vol. 44, no. 4, pp. 429--443, 2011.

\bibitem{Brisk_1997}
Deborah~J. Brisk, E.~Charles Healey, and Karen~A. Hux,
\newblock ``Clinicians’ training and confidence associated with treating school-age children who stutter,''
\newblock {\em Language, Speech, and Hearing Services in Schools}, vol. 28, no. 2, pp. 164--176, 1997.

\bibitem{Gabel_2013}
Rodney~M. Gabel,
\newblock ``School speech-language pathologists’ experiences with stuttering: an ohio survey,''
\newblock {\em eHearsay}, vol. 3, no. 1, pp. 5--23, 2013.

\bibitem{kelly_1997}
Ellen~M. Kelly, Jane~S. Martin, Kendra~E. Baker, Norma~I. Rivera, Jane~E. Bishop, Cindy~B. Krizizke, Deborah~S. Stettler, and June~M. Stealy,
\newblock ``Academic and clinical preparation and practices of school speech-language pathologists with people who stutter,''
\newblock {\em Language, Speech, and Hearing Services in Schools}, vol. 28, no. 3, pp. 195--212, 1997.

\bibitem{stLouis_1994}
Kenneth St.~Louis and C~Durrenberger,
\newblock ``What communication disorders do experienced clinicians prefer to manage?,''
\newblock {\em ASHA}, vol. 35, pp. 23--31, 35, 01 1994.

\bibitem{Coalson_2016}
Geoffrey~A. Coalson, Courtney~T. Byrd, and Elizabeth Rives,
\newblock ``Academic, clinical, and educational experiences of self-identified fluency specialists,''
\newblock {\em Perspectives of the ASHA Special Interest Groups}, vol. 1, no. 4, pp. 16--43, 2016.

\bibitem{sheikhStutterNet}
Shakeel~Ahmad Sheikh, Md~Sahidullah, Fabrice Hirsch, and Slim Ouni,
\newblock ``{StutterNet: Stuttering Detection Using Time Delay Neural Network},''
\newblock in {\em {EUSIPCO 2021 - 29th European Signal Processing Conference}}, Dublin / Virtual, Ireland, 2021.

\bibitem{Kourkounakis2021FluentNetED}
Tedd Kourkounakis, Amirhossein Hajavi, and Ali Etemad,
\newblock ``Fluentnet: End-to-end detection of stuttered speech disfluencies with deep learning,''
\newblock {\em IEEE/ACM Transactions on Audio, Speech, and Language Processing}, vol. 29, pp. 2986--2999, 2021.

\bibitem{lea_Sep28k}
Colin Lea, Vikramjit Mitra, Aparna Joshi, Sachin Kajarekar, and Jeffrey Bigham,
\newblock ``Sep-28k: A dataset for stuttering event detection from podcasts with people who stutter,''
\newblock in {\em ICASSP}, 2021.

\bibitem{librispeech}
Vassil Panayotov, Guoguo Chen, Daniel Povey, and Sanjeev Khudanpur,
\newblock ``Librispeech: An asr corpus based on public domain audio books,''
\newblock in {\em 2015 IEEE International Conference on Acoustics, Speech and Signal Processing (ICASSP)}, 2015, pp. 5206--5210.

\bibitem{yang21c_superb}
Shuwen Yang, Po-Han Chi, Yung-Sung Chuang, Cheng-I~Jeff Lai, Kushal Lakhotia, Yist~Y. Lin, Andy~T. Liu, Jiatong Shi, Xuankai Chang, Guan-Ting Lin, Tzu-Hsien Huang, Wei-Cheng Tseng, Ko~tik Lee, Da-Rong Liu, Zili Huang, Shuyan Dong, Shang-Wen Li, Shinji Watanabe, Abdelrahman Mohamed, and Hung yi~Lee,
\newblock ``{SUPERB: Speech Processing Universal PERformance Benchmark},''
\newblock in {\em Proc. Interspeech 2021}, 2021, pp. 1194--1198.

\bibitem{baevski2021wav2vec-u}
Alexei Baevski, Wei-Ning Hsu, Alexis Conneau, and Michael Auli,
\newblock ``Unsupervised speech recognition,''
\newblock {\em NeurIPS}, 2021.

\bibitem{liu2022wav2vec-u2}
Alexander~H. Liu, Wei-Ning Hsu, Michael Auli, and Alexei Baevski,
\newblock ``Towards end-to-end unsupervised speech recognition,''
\newblock in {\em 2022 IEEE Spoken Language Technology Workshop (SLT)}, 2023, pp. 221--228.

\bibitem{Peng_SSL_sv_2023}
Junyi Peng, Themos Stafylakis, Rongzhi Gu, Oldřich Plchot, Ladislav Mošner, Lukáš Burget, and Jan Černocký,
\newblock ``Parameter-efficient transfer learning of pre-trained transformer models for speaker verification using adapters,''
\newblock in {\em ICASSP 2023 - 2023 IEEE International Conference on Acoustics, Speech and Signal Processing (ICASSP)}, 2023, pp. 1--5.

\bibitem{Payal_iccasp_Siamese}
Payal Mohapatra, Bashima Islam, Md~Tamzeed Islam, Ruochen Jiao, and Qi~Zhu,
\newblock ``Efficient stuttering event detection using siamese networks,''
\newblock in {\em ICASSP 2023 - 2023 IEEE International Conference on Acoustics, Speech and Signal Processing (ICASSP)}, 2023, pp. 1--5.

\bibitem{bayerl22b_interspeech}
Sebastian~Peter Bayerl, Dominik Wagner, Elmar Noeth, and Korbinian Riedhammer,
\newblock ``{Detecting Dysfluencies in Stuttering Therapy Using wav2vec 2.0},''
\newblock in {\em Proc. Interspeech 2022}, 2022, pp. 2868--2872.

\bibitem{bayerl23_interspeech}
Sebastian~P. Bayerl, Dominik Wagner, Ilja Baumann, Florian Hönig, Tobias Bocklet, Elmar Nöth, and Korbinian Riedhammer,
\newblock ``{A Stutter Seldom Comes Alone – Cross-Corpus Stuttering Detection as a Multi-label Problem},''
\newblock in {\em Proc. INTERSPEECH 2023}, 2023, pp. 1538--1542.

\bibitem{wav2vec2}
Alexei Baevski, Henry Zhou, Abdelrahman Mohamed, and Michael Auli,
\newblock ``wav2vec 2.0: a framework for self-supervised learning of speech representations,''
\newblock in {\em Proceedings of the 34th International Conference on Neural Information Processing Systems}, Red Hook, NY, USA, 2020, NIPS '20, Curran Associates Inc.

\bibitem{harvill22_interspeech}
John Harvill, Mark Hasegawa-Johnson, and Chang~D. Yoo,
\newblock ``{Frame-Level Stutter Detection},''
\newblock in {\em Proc. Interspeech 2022}, 2022, pp. 2843--2847.

\bibitem{Chen2021WavLMLS}
Sanyuan Chen, Chengyi Wang, Zhengyang Chen, Yu~Wu, Shujie Liu, Zhuo Chen, Jinyu Li, Naoyuki Kanda, Takuya Yoshioka, Xiong Xiao, Jian Wu, Long Zhou, Shuo Ren, Yanmin Qian, Yao Qian, Micheal Zeng, and Furu Wei,
\newblock ``Wavlm: Large-scale self-supervised pre-training for full stack speech processing,''
\newblock {\em IEEE Journal of Selected Topics in Signal Processing}, vol. 16, pp. 1505--1518, 2021.

\bibitem{Howell_uclass}
Peter Howell, Stephen Davis, and Jon Bartrip,
\newblock ``The university college london archive of stuttered speech (uclass),''
\newblock {\em Journal of speech, language, and hearing research : JSLHR}, vol. 52, pp. 556--69, 05 2009.

\bibitem{bayerl_KSoFKasselState_2022}
Sebastian Bayerl, Alexander Wolff~von Gudenberg, Florian H{\"o}nig, Elmar Noeth, and Korbinian Riedhammer,
\newblock ``Ksof: The kassel state of fluency dataset -- a therapy centered dataset of stuttering,''
\newblock in {\em Proceedings of the Language Resources and Evaluation Conference}, Marseille, France, June 2022, pp. 1780--1787, European Language Resources Association.

\bibitem{Farhad_Dysarthria}
Farhad Javanmardi, Saska Tirronen, Manila Kodali, Sudarsana~Reddy Kadiri, and Paavo Alku,
\newblock ``Wav2vec-based detection and severity level classification of dysarthria from speech,''
\newblock in {\em ICASSP 2023 - 2023 IEEE International Conference on Acoustics, Speech and Signal Processing (ICASSP)}, 2023, pp. 1--5.

\bibitem{Lian2023UnconstrainedDM}
Jiachen Lian, Carly Feng, Naasir Farooqi, Steve Li, Anshul Kashyap, Cheol~Jun Cho, Peter Wu, Robin Netzorg, Tingle Li, and Gopala~Krishna Anumanchipalli,
\newblock ``Unconstrained dysfluency modeling for dysfluent speech transcription and detection,''
\newblock {\em 2023 IEEE Automatic Speech Recognition and Understanding Workshop (ASRU)}, pp. 1--8, 2023.

\bibitem{Lian2024TowardsHS}
Jiachen Lian and Gopala Anumanchipalli,
\newblock ``Towards hierarchical spoken language disfluency modeling,''
\newblock in {\em Proceedings of the 18th Conference of the European Chapter of the Association for Computational Linguistics (Volume 1: Long Papers)}, St. Julian{'}s, Malta, Mar. 2024, pp. 539--551, Association for Computational Linguistics.

\bibitem{shih_interface}
Yi-Jen Shih and David Harwath,
\newblock ``{Interface Design for Self-Supervised Speech Models},''
\newblock in {\em Proc. INTERSPEECH 2024}, 2024.

\bibitem{Macwhinney_2000}
Brian Macwhinney,
\newblock ``The childes project: Tools for analyzing talk: Transcription format and programs (3rd ed.).,''
\newblock {\em Lawrence Erlbaum Associates Publishers}, vol. 1, 2000.

\bibitem{Ratner_Brundage_2020}
Nan~Bernstein Ratner and Shelley~B. Brundage,
\newblock ``A clinician’s complete guide to clan and praat,''
\newblock 2020.

\bibitem{Ratner_2018}
Nan {Bernstein Ratner} and Brian MacWhinney,
\newblock ``Fluency bank: A new resource for fluency research and practice,''
\newblock {\em Journal of Fluency Disorders}, vol. 56, pp. 69--80, 2018.

\bibitem{pmlr-v202-radford23a_whisper}
Alec Radford, Jong~Wook Kim, Tao Xu, Greg Brockman, Christine Mcleavey, and Ilya Sutskever,
\newblock ``Robust speech recognition via large-scale weak supervision,''
\newblock in {\em Proceedings of the 40th International Conference on Machine Learning}. 23--29 Jul 2023, vol. 202 of {\em Proceedings of Machine Learning Research}, pp. 28492--28518, PMLR.

\bibitem{baevski22a_data2vec}
Alexei Baevski, Wei-Ning Hsu, Qiantong Xu, Arun Babu, Jiatao Gu, and Michael Auli,
\newblock ``data2vec: A general framework for self-supervised learning in speech, vision and language,''
\newblock in {\em Proceedings of the 39th International Conference on Machine Learning}, Kamalika Chaudhuri, Stefanie Jegelka, Le~Song, Csaba Szepesvari, Gang Niu, and Sivan Sabato, Eds. 17--23 Jul 2022, vol. 162 of {\em Proceedings of Machine Learning Research}, pp. 1298--1312, PMLR.

\end{thebibliography}

\end{document}